\documentclass[preprint,12pt]{elsarticle}

\usepackage{float}
\usepackage{xcolor}
\usepackage{marvosym}

\usepackage{amssymb}
\usepackage{amsmath}
\usepackage{float}
\usepackage{multirow} 
\usepackage{url}

\journal{Medical Image Analysis}

\begin{document}

\begin{frontmatter}



\title{AGFS-Tractometry: A Novel Atlas-Guided Fine-Scale Tractometry Approach for Enhanced Along-Tract Group Statistical Comparison Using Diffusion MRI Tractography}


\cortext[cor1]{Corresponding author: Fan Zhang, fan.zhang@uestc.edu.cn}
\cortext[cor2]{Fan Zhang and Lauren O'Donnell are co-senior-authors.}
\author[a]{Ruixi Zheng}
\author[a]{Wei Zhang}
\author[a]{Yijie Li}
\author[a]{Xi Zhu}
\author[b]{Zhou Lan}
\author[b,c]{Jarrett Rushmore}
\author[b]{Yogesh Rathi}
\author[b]{Nikos Makris}
\author[b]{Lauren J. O'Donnell\corref{cor2}}
\author[a]{Fan Zhang\corref{cor1},\corref{cor2}}

\affiliation[a]{organization={University of Electronic Science and Technology of China},
            addressline={No. 2006, Xiyuan Ave}, 
            city={Chengdu},
            postcode={611731}, 
            state={Sichuan},
            country={China}}

\affiliation[b]{organization={Brigham and Women’s Hospital, Harvard Medical School},
            addressline={75 Francis St}, 
            city={Boston},
            postcode={02215}, 
            state={MA},
            country={USA}}

\affiliation[c]{organization={Boston University},
            addressline={72 E Concord St}, 
            city={Boston},
            postcode={02118}, 
            state={MA},
            country={USA}}

\begin{abstract}
Diffusion MRI (dMRI) tractography is currently the only method for in vivo mapping of the brain’s white matter (WM) connections. Tractometry is an advanced tractography analysis technique for along-tract profiling to investigate the morphology and microstructural properties along the fiber tracts. Tractometry has become an essential tool for studying local along-tract differences between different populations (e.g., health vs disease). In this study, we propose a novel atlas-guided fine-scale tractometry method, namely AGFS-Tractometry, that leverages tract spatial information and permutation testing to enhance the along-tract statistical analysis between populations. There are two major contributions in AGFS-Tractometry. First, we create a novel atlas-guided tract profiling template that enables consistent, fine-scale, along-tract parcellation of subject-specific fiber tracts. Second, we propose a novel nonparametric permutation testing group comparison method to enable simultaneous analysis across all along-tract parcels while correcting for multiple comparisons. We perform experimental evaluations on synthetic datasets with known group differences and in vivo real data. We compare AGFS-Tractometry with two state-of-the-art tractometry methods, including Automated Fiber-tract Quantification (AFQ) and BUndle ANalytics (BUAN). Our results show that the proposed AGFS-Tractometry obtains enhanced sensitivity and specificity in detecting local WM differences. In the real data analysis experiments, AGFS-Tractometry can identify more regions with significant differences, which are anatomically consistent with the existing literature. Overall, these demonstrate the ability of AGFS-Tractometry to detect subtle or spatially localized WM group-level differences. The created tract profiling template and related code are available at: \url{https://github.com/ZhengRuixi/AGFS-Tractometry.git}.
\end{abstract}




\end{frontmatter}



\section{Introduction}
\label{sec1}
\begin{figure}[!t]
\centering
\includegraphics[width=13.5cm,height=10cm]{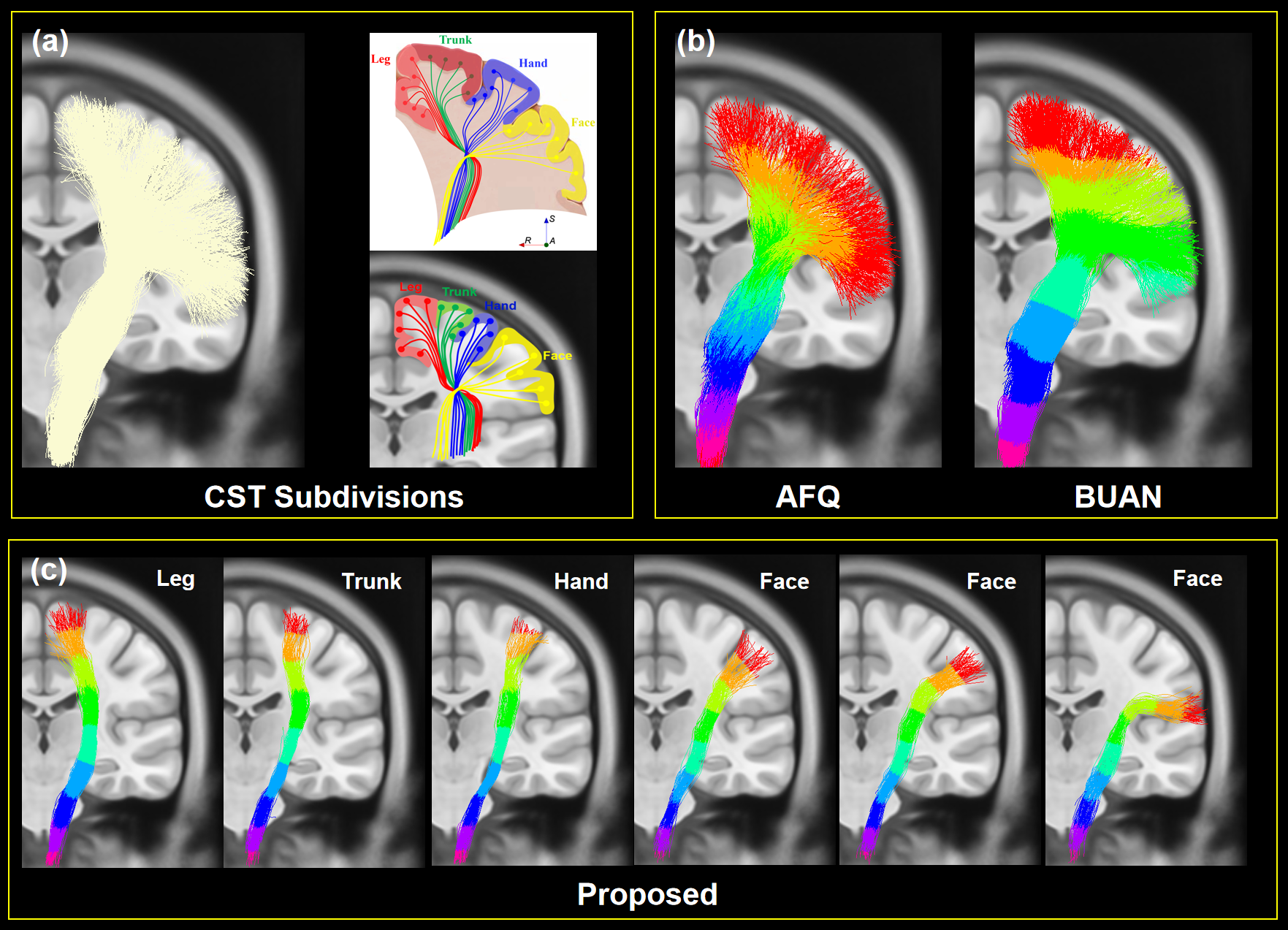}
\caption{Different tract profiling strategies in tractometry, including AFQ, BUAN, and AGFS-Tractometry. (a) Illustration of the CST with color-coded clusters indicating anatomical subdivisions. (b) Tract profiling along the entire fiber tract using the traditional AFQ and BUAN methods, with results shown as color-coded parcels. (c) Our proposed AGFS-Tractometry method can achieve finer subdivisions, distinguishing different motor pathways for the leg, trunk, hand, and face.}\label{fig1}
\end{figure}

Diffusion magnetic resonance imaging (dMRI) is currently the only technique for \textit{in vivo} reconstruction of the brain’s white matter (WM) connections, providing a noninvasive opportunity to characterize their underlying tissue microstructure by probing water molecule diffusion and mapping fiber tracts \cite{basser2000}. In tractography, fiber tracts are modeled as collections of streamlines that reflect the pathways of WM fibers traversing various anatomical regions. For example, Figure \ref{fig1}a gives a visualization of the corticospinal tract(CST\footnote{In line with common practice in the tractography field, the term "CST" is used throughout to describe the major descending pathway from the motor cortex. We acknowledge that the CST is a subpart of the PyT, which also includes several other fibers projecting to deep GM structures\cite{Kjer2025}.}, i.e., its trajectory within the pyramidal tract/PyT) \cite{Kjer2025} and its somatotopic organization, with four major subdivisions connecting to different cortical regions. Tractometry is an advanced tractography analysis technique for along-tract analysis to investigate the morphology and microstructural properties along anatomical fiber tracts \cite{colby2012a,corouge2006,yeatman2012,chandio2020,odonnell2009}. Unlike traditional approaches that compute a summary statistic of the entire tract (e.g., streamline count or volume), tractometry enables the study of local regions along the fiber tract and thus can be more sensitive to detect subtle WM group differences for between-population comparisons \cite{zhang2022}. Over the past decade, group comparison of the WM fiber tracts across populations using tractometry has increasingly contributed to our understanding of normal brain development and assisted the identification of potential biomarkers for neurological disorders \cite{colby2012a,chamberland2021,neher2024}.

In general, the first step of performing tractometric analyses is \textit{tract profiling}, which describes the distribution of the WM measures along the fiber tract. In this step, the fiber tract needs to be segmented into smaller parcels, from which the WM measure of interest is extracted (e.g., average fractional anisotropy, FA). Currently, state-of-the-art methods for tract profiling are the AFQ \cite{yeatman2012} and BUAN \cite{chandio2020} methods. In AFQ, streamlines are resampled to $n$ equidistant points along their arc length (e.g., $n=100$), and each tract parcel is determined by its position indices of points along the streamline  (Figure \ref{fig1}b, left). BUAN proposes a more advanced parcellation approach to assign each streamline point to the closest point on a tract-centerline composed of n points  (Figure \ref{fig1}b, right). Despite the widespread use of these two methods, a limitation is their focus on the entire fiber tract, overlooking possible anatomical subdivisions within the tract. For example, the CST pathway has a somatotopic organization with multiple subdivisions connecting to the leg, trunk, hand, and face cortical areas \cite{he2023} (see Figure \ref{fig1}a). The existing AFQ and BUAN methods are not effective in distinguishing tract parcels belonging to the different anatomical subdivisions. Therefore, we propose a novel tract profiling strategy that leverages what we term a fine-scale along-tract parcellation. This fine-scale approach provides a higher spatial sampling resolution by parcellating the tract not only along its length, as in AFQ and BUAN, but also across its diameter by leveraging fiber cluster subdivisions. This enables a detailed investigation of local tract regions (Figure \ref{fig1}c) and and yields a substantially greater number of parcels than conventional approaches (e.g., increasing from 100 parcels along the tract in AFQ to 100 parcels per cluster × 15 clusters = 1500 parcels in total).

After creating the tract profile, the next step in tractometry is \textit{statistical comparison}, which aims to identify local tract regions with statistically significant group differences. Current tractometry methods typically apply hypothesis testing at each parcel along the tract profile. For example, AFQ uses Student’s t-test at each node of the tract profile, whereas BUAN employs a linear mixed model (LMM) to account for both fixed and random effects in the experimental data. In both cases, multiple comparisons across multiple tract parcels are controlled using methods such as the false discovery rate (FDR) and permutation test \cite{benjamini1995}. The large number of tract parcels in tractometric analyses, especially in fine-scale profiling, increases the number of statistical tests, which in turn demands stricter correction for multiple comparisons. As a result, the aforementioned conventional statistical analysis methods may lack the sensitivity required to reliably detect group differences. This limitation is a well-recognized challenge in voxel-based neuroimaging approaches such as tract-based spatial statistics (TBSS) \cite{smith2006} and voxel-based morphometry (VBM) \cite{ashburner2000}, which involve conducting a large number of statistical tests across individual voxels and require correction for multiple comparisons. Currently, clustering-thresholding techniques are widely used in neuroimaging to enhance statistical sensitivity by leveraging spatial contiguity, particularly in voxel-based analysis (VBA) \cite{nichols2002,woo2014}. In tractography, similar approaches have been applied to detect tract-level WM differences \cite{zhang2018a,wu2018} or structural network changes \cite{zalesky2012}. In the context of tractometry, cluster-wise correction has been explored in approaches such as correlational tractography \cite{Yeh2025} and Tracula \cite{Yendiki2011}, which leverage spatial clustering strategies to improve sensitivity. However, no prior work has applied cluster-thresholding in tractometry, where multiple along-tract comparisons pose a statistical challenge. In our study, we introduce a novel tractometry framework for anatomically guided fiber profiling and propose a clustering-thresholding-based statistical method to enable simultaneous analysis across all along-tract parcels while correcting for multiple comparisons.

In this paper, we design a novel atlas-guided fine-scale tractometry method, namely \textit{AGFS-Tractometry}, that leverages tract spatial information and permutation testing to enhance the along-tract statistical analysis between populations. Our study has two major contributions, corresponding to the two major steps in tractometric analyses. First, for tract profiling, we introduce a novel approach that begins with the creation of a population-wise tract profiling template. This template is constructed based on an anatomically curated WM fiber tract atlas (i.e., the ORG-atlas) \cite{zhang2018b,zhang2019}. In this atlas, major tracts are subdivided into multiple fiber clusters. Our method leverages these clusters to achieve a fine-scale parcellation, which we define as a higher spatial sampling resolution achieved by parcellating not only along the tract's length but also across its diameter (as illustrated in Figure 1). Furthermore, the created template can be applied to individual tractography data to extract subject-specific tract profiles. In this way, using the template acts as a common space where all subjects are mapped into a unified framework. This naturally establishes correspondence for local along-tract parcels across multiple subjects, facilitating robust group comparisons. Second, for group-wise statistical comparison, we introduce a novel nonparametric, permutation testing method to enable simultaneous analysis across all along-tract parcels while correcting for multiple comparisons. The design of this method is inspired by voxel-based cluster-thresholding approaches, which leverage spatial neighborhoods to strengthen statistical sensitivity \cite{nichols2002}. In our approach, we develop a parcel spatial neighborhood construction strategy to form spatially adjacent parcels into communities. Then the parcels within communities can mutually reinforce each other to enhance the detection of statistically significant differences. We evaluate our AGFS-Tractometry by comparing it with two widely used tractometry techniques, AFQ and BUAN. First, we conduct experiments on synthetic datasets that include ground truth, meaning the group differences along the fiber tracts are known. Our approach exhibits enhanced sensitivity and specificity in detecting these ground truth differences. Additionally, we assess our method in two real-world scenarios: one focuses on investigating sex differences in WM fiber tracts, and the other on examining the effects of disease on WM in autism. The results demonstrate that our method achieves improved sensitivity in identifying localized differences.

\section{Method}
\label{sec2}

\begin{figure}[!t]
\centering
\includegraphics[width=13.5cm,height=8.5cm]{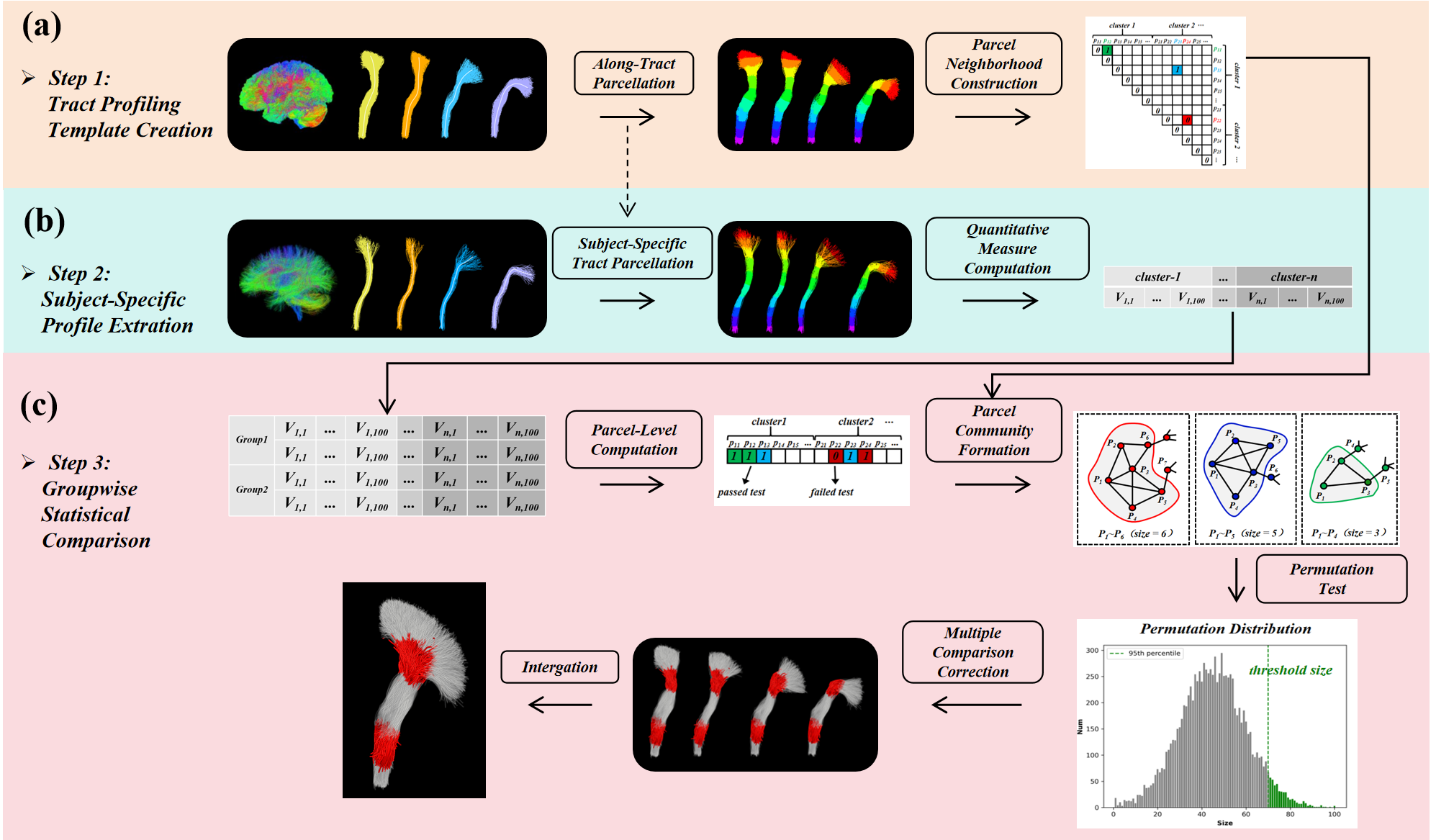}
\caption{Method Overview. (a) Step 1: Tract Profiling Template Creation. Along-tract parcellation is guided by centerlines computed from the ORG-atlas, followed by neighborhood construction to generate a parcel neighborhood matrix; (b) Step 2: Subject-Specific Profile Extraction. Subject-specific tract parcellation is performed, and quantitative measures are computed for each parcel to generate individual tract profiles; (c) Step 3: Groupwise Statistical Comparison. Group-level comparisons are conducted at the parcel level. Statistically significant parcels are identified and grouped into communities based on the parcel neighborhood matrix. A permutation test determines the significance of each community according to size thresholds, resulting in the final detection of group differences.}\label{fig2}
\end{figure}

\subsection{Overview}
\label{subsec21}
The goal of AGFS-Tractometry is to identify local regions along the tract of interest (e.g., CST) that exhibit statistically significant differences when comparing two populations. Figure \ref{fig2} gives an overview of the method, including three major steps. First, we create an atlas-guided fine-scale tract profiling template. This template defines the parcellation of fiber tracts into multiple parcels along their length (Section \ref{subsubsec221}) and the spatial neighborhood relationships between these parcels (Section \ref{subsubsec222}). Second, we compute subject-specific tract profiles by applying the profiling template to subject tractography data. This step identifies subject-specific along-tract parcels (Section \ref{subsubsec231}) and extracts parcel quantitative measures (Section \ref{subsubsec232}). Third, we perform statistical comparisons of the subject-specific tract profiles between the two groups. In this step, a parcel-level statistical test is performed to identify uncorrected group differences (an uncorrected p-value) for each parcel (Section \ref{subsubsec241}). Subsequently, a nonparametric cluster-thresholding-based permutation test, relying on summary statistics derived from parcel communities formed according to the parcel spatial neighborhood relationships (Section \ref{subsubsec242}), is performed to correct for multiple comparisons (Section \ref{subsubsec243}).

\subsection{Atlas-Guided Fine-Scale Tract Profiling Template Creation}
\label{subsec22}
The proposed tract profiling template is developed based on an anatomically curated fiber clustering atlas, i.e., the ORG atlas. We extend the ORG atlas to further subdivide fiber tracts into along-tract parcels and establish spatial neighborhood relationships among these parcels. 

\subsubsection{Fine-Scale Along-Tract Parcellation}
\label{subsubsec221}
The ORG atlas was generated by creating dense tractography maps of 100 individual subjects and then applying fiber clustering to group fiber streamlines across subjects according to their similarity in shape and location \cite{zhang2018b,zhang2019}. It includes an anatomical tract parcellation into 58 major WM fiber tracts, including well-known fiber tracts such as the arcuate fasciculus (AF) and the CST. Each anatomical fiber tract is composed of multiple fiber clusters that represent subdivisions of the entire tract (see Figure \ref{fig2}a). Building on this cluster-based subdivision, we propose to further parcellate each cluster along its length and thus obtain a fine-scale along-tract parcellation.

Specifically, in the ORG atlas, a fiber cluster is represented as a collection of streamlines, where each streamline consists of multiple points along its length. For each cluster, we adopt the centerline profiling strategy proposed in BUAN for along-cluster parcellation. A centerline is first generated by resampling each fiber streamline to n equidistant points and computing the average coordinates of the corresponding points across all streamlines at each location. Next, the spatial distances between each streamline point in the cluster and each centerline point are calculated, followed by assigning each streamline point to its nearest centerline point. This process results in an along-cluster parcellation of each fiber cluster into n parcels. One advantage of performing cluster-level centerline profiling is that in the ORG atlas, fiber clusters are created by grouping streamlines that share similar geometric trajectories and lengths. In this way, streamline points within the same parcel maintain comparable distances to the centerline point, ensuring that each parcel reflects the local geometry of fiber clusters. This can avoid a potential issue observed in BUAN,  where imperfect alignment at tract endpoint regions may lead to the misclassification of points from the same location into different parcels (see Figure \ref{fig1}b). Lastly, for a specific anatomical tract composed of $m$ clusters, a fine-scale along-tract parcellation into $m$ × $n$ parcels is obtained to serve as the template for tract profiling.

\subsubsection{Parcel Neighborhood Construction}
\label{subsubsec222}
After obtaining the along-tract parcellation template, we construct a parcel neighborhood to identify nearby template parcels with similar WM anatomy. This is essential for combining parcels into communities for summary statistics computation (Section \ref{subsec24}). In contrast to the standard along-tract parcellation methods used in AFQ and BUAN, where neighboring parcels are determined by their positional indices (e.g., parcels at positions $k$ and $k+1$ are considered neighbors), our parcellation approach incorporates not only intra-cluster but also inter-cluster parcels. This introduces complexity in defining parcel neighborhoods, as the relationships between parcels belonging to different clusters are not as straightforward.

\begin{figure}[!t]
\centering
\includegraphics[width=14cm,height=5.5cm]{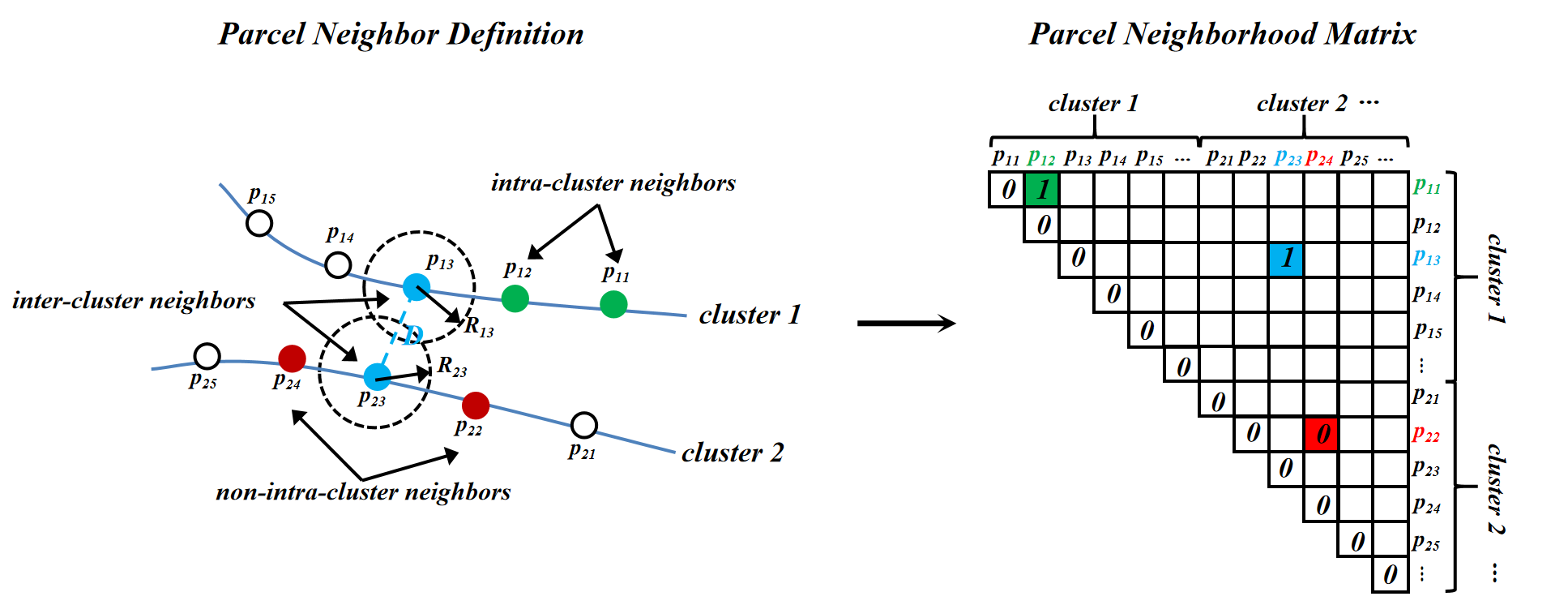}
\caption{Illustration of the parcel neighborhood construction process: (a) Intra- and inter-cluster parcel neighbor relationship, (b) parcel neighborhood matrix.}\label{fig3}
\end{figure}

Our method proposes the following strategy for parcel neighborhood construction. There are two distinct positional relationships between parcels along a tract: (1) intra-cluster parcels that belong to the same cluster, and (2) inter-cluster parcels that belong to the same tract but are located in different clusters (see Figure \ref{fig3}). For intra-cluster parcels, it is straightforward that two parcels are neighbors if they are adjacent. For inter-cluster WM parcels, we consider two parcels neighbors if they are spatially close, following the condition below: 

\begin{equation}
  N(i, j) = 
  \begin{cases}
    1, & D(i, j) < R(i) + R(j) \\
    0, & \text{otherwise}
  \end{cases}
\end{equation}
where $D(i, j)$ is the Euclidean distance between the centerline points of parcels $p_i$ and $p_j$, and $R(\cdot)$ is the radius of each parcel defined as the average of the distances between all points assigned to the parcel and the parcel centerline point. Note that the radius $R$ is thus defined in a parcel-specific manner to adapt to parcel-specific geometry. $N(i, j)$ represents the neighbor relationship between parcels $p_i$ and $p_j$, with a value of 1 indicating the presence of a neighbor relationship.

One key advantage of constructing the parcel neighborhood is that it preserves the structural continuity within a tract by establishing spatial relationships between adjacent parcels. This helps mitigate the risk of information fragmentation that can arise from overly fine-scale subdivisions in whole-brain WM parcellation, facilitating a more comprehensive investigation of structural patterns.

\subsection{Subject-Specific Tract Profile Extraction}
\label{subsec23}
We extract subject-specific tract profiles for each subject by applying the created tract profiling template to subject tractography data. This includes the identification of the tract of interest and its along-tract parcellation, followed by the computation of the quantitative measures of each parcel. 

\subsubsection{Subject-Specific Along-Tract Parcellation}
\label{subsubsec231}
For each subject, whole-brain tractography (see Section \ref{subsubsec312}) is first aligned to the ORG atlas space using a tractography-based registration method \cite{odonnell2012}. A fiber spectral embedding is then performed to compute the similarity between the subject's fibers and those in the atlas, followed by assigning each of the subject's fibers to the corresponding atlas cluster \cite{odonnell2007}. Based on the tract annotations in the ORG atlas, subject-specific fiber tracts and their associated clusters can subsequently be identified.

Subject-specific along-tract parcellation is then performed by subdividing the subject fiber clusters according to the profiling template. This is done by calculating the distance between the subject's fiber points and the template centerline points and assigning each fiber point to its nearest centerline point. In this manner, each subject-specific tract is parcellated based on our tract profiling template, ensuring that the parcels from different subjects correspond to one another by using the template as an intermediary.

\subsubsection{Quantitative Measure Computation}
\label{subsubsec232}
Quantitative measures for each parcel are extracted to describe the along-tract distribution of WM microstructure for subject-specific tract profiling. In our study, we utilize the unscented Kalman filter (UKF) \cite{reddy2016,vos2013,farquharson2013} tractography method for fiber tracking, which enables the direct computation of fiber-specific microstructure measures (e.g., FA) at points along each fiber streamline. Unlike standard processing strategies that map voxel-wise microstructural measures to fiber streamlines, UKF allows for the estimation of point-specific microstructural properties (See Section \ref{subsubsec312} for details). Leveraging this unique advantage, we adopt a distance-weighted strategy \cite{yeatman2012}, in which fiber points closer to the centerline contribute more significantly to the measurement, as described below:  
\begin{equation}
  D_M(\mathbf{x}, \boldsymbol{\mu}) = \sqrt{(\mathbf{x} - \boldsymbol{\mu})^T \boldsymbol{\Sigma}^{-1} (\mathbf{x} - \boldsymbol{\mu})} \tag{2}
\end{equation}
\begin{equation}
    \text{Metric}_{\text{weighted}} = \frac{D_M(\mathbf{x}_i, \boldsymbol{\mu}) \cdot \text{Metric}_i}{\sum_{i=1}^n D_M(\mathbf{x}_i, \boldsymbol{\mu})} \tag{3}
\end{equation}
where $D_M(\mathbf{x}, \boldsymbol{\mu})$ represents the Mahalanobis distance between a streamline point $\mathbf{x}$ and the mean $\boldsymbol{\mu}$ of a distribution with covariance matrix $\boldsymbol{\Sigma}$ computed by all points within the parcel. $\mathrm{Metric}_i$ is the diffusion metric for the $i$-th point (e.g., FA). The parcel-level diffusion metric, $\mathrm{Metric}_{\mathrm{weighted}}$, is then computed as a weighted average, where the weights are derived from the Mahalanobis distance $D_M(\mathbf{x}, \boldsymbol{\mu})$. In our study, the diffusion measure FA is used, while we note that any other measures of interest (e.g., mean diffusivity (MD) and number of fiber points) can be used for quantitative parcel measures to compute the subject-specific tract profile. Furthermore, it is important to note that while the UKF is employed in this study for fiber tracking, our proposed framework is not limited to this specific method. The framework is designed to be generalizable and can accommodate any fiber tracking algorithm that provides point-specific microstructural measures along fiber streamlines. This flexibility allows for the integration of various fiber tracking techniques, ensuring that our method can be adapted to different research contexts and data types.

\subsection{Groupwise Tract Profile Statistical Comparison}
\label{subsec24}
\begin{figure}[!t]
\centering
\includegraphics[width=13.5cm,height=9cm]{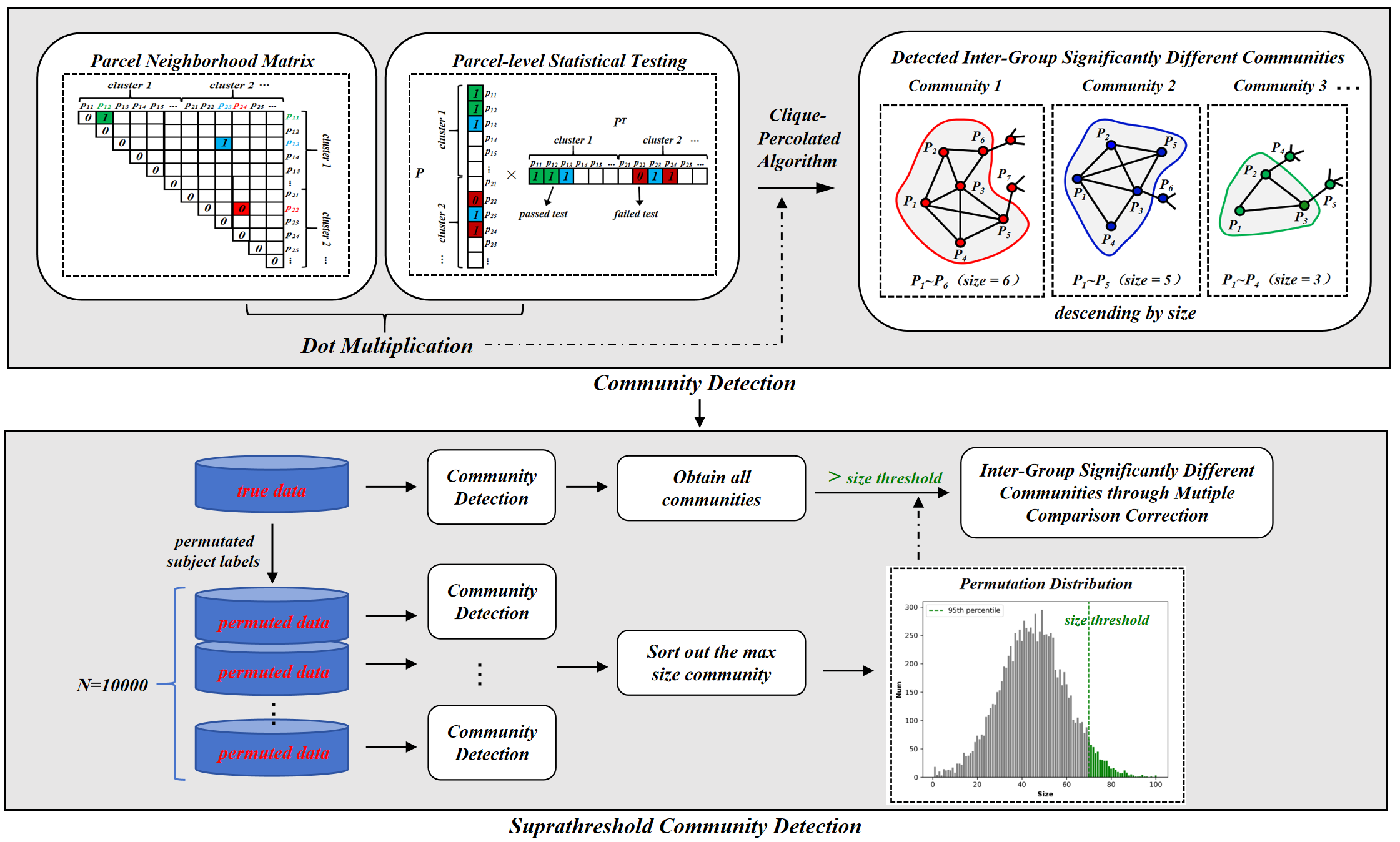}
\caption{Method for Community Detection and Statistical Validation. The upper panel illustrates the community detection process. A parcel neighborhood matrix and a parcel-level statistical testing matrix are constructed and combined via dot multiplication to form a parcel graph. The clique percolation (CP) algorithm is then applied to detect parcel communities based on the graph structure. Communities are sorted by size in descending order. The lower panel depicts the suprathreshold community detection. A null distribution of maximum community sizes is generated by repeatedly permuting subject labels and applying community detection on the permuted datasets. Communities obtained from the true data are considered significant if their size exceeds the threshold derived from the null distribution (e.g., the 95th percentile), ensuring control of false positive rates.}\label{fig4}
\end{figure}

After generating the tract profile for each subject, statistical comparisons are conducted between the two groups to identify local along-tract regions exhibiting significant differences. To address multiple comparisons across a large number of parcels, we leverage cluster-thresholding-based permutation testing to enhance statistical group difference analysis while correcting for multiple comparisons (see Figure \ref{fig4}). 

\subsubsection{Parcel-Level Statistical Comparison}
\label{subsubsec241}
A parcel-level statistical test is performed for each along-tract parcel to identify the parcels with uncorrected group differences at a predetermined primary threshold. These parcels are referred to as the suprathreshold parcels. Specifically, we perform a null hypothesis test that computes group differences for each parcel to derive an uncorrected p-value. Commonly used statistical test methods, such as Student’s t-test and generalized linear model (GLM), can be used. We set the primary threshold to \textit{p}-value$=0.05$, i.e., the parcels with uncorrected \textit{p}-values $\leq 0.05$ pass the test.

\subsubsection{Suprathreshold Parcel Community Formation}
\label{subsubsec242}
A suprathreshold parcel community is defined as a set of multiple neighboring suprathreshold parcels that pass the parcel-level test. Given the defined parcel neighborhood graph (Section \ref{subsubsec222}) and the parcels that passed the parcel-level statistic test (Section \ref{subsubsec231}), we apply clique percolation (CP) \cite{palla2005} to identify the suprathreshold parcel communities. In graph theory, CP is a well-established method for detecting densely connected node parcels based on known edges for the parcel neighborhood graph. It has been successfully applied across various domains, such as analyzing structural and functional brain connectivity \cite{sporns2016}, community detection in protein networks \cite{jonsson2006}, studying social networks \cite{gonzalez2007}, and whole brain tractography comparison \cite{zhang2018a}. In brief, given the neighborhood relationships among all nodes on a graph, CP first detects all cliques, i.e. fully connected subgraphs. In our context, nodes represent WM parcels, and edges represent neighborhood relationships between parcels. A clique with $k$ nodes is called a $k$-clique and two cliques are adjacent if they share at least $h$ nodes. Then, a node community is calculated as the maximal union of cliques that could be reached from each other through a series of adjacent cliques.

Using the parcel neighborhood graph derived from all along-tract parcels (as computed using the ORG atlas prior to statistical analysis; see Section \ref{subsubsec221}), we first construct two $n$×$n$ binary matrices: (1) a parcel neighborhood matrix indicating whether each pair of parcels are spatially adjacent (Section \ref{subsubsec222}), and (2) a parcel-level significance matrix, where an element is 1 only if a pair of parcels are both suprathreshold in the statistical testing (Section \ref{subsubsec241}). We then perform an element-wise matrix multiplication between the neighborhood matrix and the parcel-level significance matrix. This operation retains only the neighborhood graph edges connecting suprathreshold parcels, effectively removing edges between non-significant nodes, and thus retaining subgraphs of connected suprathreshold parcels. Based on the retained subgraphs, CP is applied to identify all cliques and then integrate the adjacent parcel cliques as communities, as illustrated in Figure \ref{fig4}. In our experiments, we set $k$=3, $h$=2 that defined a pair of 3-cliques ($k$ = 3) as adjacent if there were at least two shared parcels ($h$ = 2). We note that the parameter selections of $k$=3 and $h$=2 are the minimum values that allow for CP \cite{palla2005}, ensuring all possible suprathreshold parcel communities are identified.

\subsubsection{Nonparametric Permutation Test For Multiple Comparison Correction}
\label{subsubsec243}
We use a nonparametric permutation test to evaluate whether a suprathreshold parcel community is statistically different between groups. This test is chosen for its conceptual simplicity, minimal reliance on statistical assumptions, and its suitability for suprathreshold analyses that require correction for multiple comparisons across numerous brain WM parcels \cite{nichols2002,holmes1996}. Prior research has successfully applied permutation tests to detect group-level WM differences along single-fiber tracts \cite{odonnell2009,taquet2012,wang2016}.

In our proposed method, we apply the permutation test to single-tract analyses. For each community (defined as a group of WM parcels), we determine a corrected significance level based on its size (i.e., the number of parcels it contains). Similar to the suprathreshold fiber cluster test in \cite{zhang2018a}, we use the maximal suprathreshold community size as the summary statistic. During each permutation run, this maximal community size is calculated to build a null distribution, which enables control of the family-wise error rate for false positives.

Figure~\ref{fig4} illustrates the full permutation testing process. After performing parcel-level statistical comparisons (Section~\ref{subsubsec231}), we extract all communities identified in the observed data using CP (Section~\ref{subsubsec232}). To control for false positives, we randomly shuffle subject labels and record the maximal suprathreshold community size across multiple permutations. As shown in Figure~\ref{fig4}b, this generates a null distribution of maximal suprathreshold community sizes that serves as a reference for multiple comparison correction. We then apply a statistical threshold, typically the 95th percentile of this null distribution ($p < 0.05$), to the suprathreshold communities in the correctly labeled data. Only those communities exceeding this size threshold are retained as significant.

\section{Experiments and Results}
\label{sec3}

\subsection{Evaluation Datasets}
\label{subsec31}
We evaluate the proposed method using two dMRI datasets from different sources: the Human Connectome Project Young Adult (HCP-YA) dataset \cite{vanessen2013} and the Autism Brain Imaging Data Exchange II database contributed by New York University (ABIDE II-NYU)  \cite{dimartino2017}. The HCP-YA dataset comprised dMRI data from young healthy adults (22-37 years), and the ABIDE II-NYU dataset included dMRI data from children diagnosed with Autism Spectrum Disorders (ASD) and those who are typically developing. In our study, we include 100 participants (29.1±3.7 years; 54F, 46M) from the HCP-YA dataset to examine gender differences in brain WM and 52 participants from the ABIDE II-NYU dataset (ASD:30, 10.5±6.8 years; TC:22,  9.9±3.6 years; two-tailed $t$-test, p-value=0.69) to study WM defects in ASD.

\subsubsection{MRI Acquisition and Preprocessing}
\label{subsubsec311}
The dMRI data of the above participants are publicly available in the HCP-YA \cite{wang2016} and ABIDE II-NYU \cite{vanessen2013} datasets. The detailed acquisition parameters are described in \cite{wang2016} and \cite{vanessen2013}. Briefly, the dMRI data in HCP-YA is acquired on a customized 3T Connectome Siemens Skyra scanner, with the following acquisition parameters: TE = 89.5\,ms, TR = 5520\,ms, voxel size = $1.25 \times 1.25 \times 1.25\,\mathrm{mm}^3$, and a total of 288 volumes including 18 baseline images and 90 diffusion-weighted images at each of the three shells of $b = 1000/2000/3000\,\mathrm{s}/\mathrm{mm}^2$. We use the preprocessed dMRI data provided in the HCP-YA database. The dMRI data in ABIDE II-NYU was acquired on a 3T Siemens Allegra MRI scanner with the following acquisition parameters: TE = 78\,ms, TR = 5200\,ms, voxel size = $3 \times 3 \times 3\,\mathrm{mm}^3$, 1 baseline image at $b = 0\,\mathrm{s}/\mathrm{mm}^2$, 64 diffusion-weighted images at $b = 1000\,\mathrm{s}/\mathrm{mm}^2$. We preprocess the ABIDE II-NYU data using our well-established dMRI processing pipeline \cite{glasser2013}, including brain masking, eddy current-induced correction, motion correction, and EPI distortion correction.

\subsubsection{Whole-Brain Tractography}
\label{subsubsec312}
Whole-brain tractography is performed using the UKF tractography method \cite{reddy2016,vos2013,farquharson2013}, as implemented in the ukftractography package via SlicerDMRI \cite{Norton2017-gc,Zhang2020-ce}. The UKF approach fits a two-tensor mixture model to the dMRI data while tracking fibers, incorporating prior information from the previous step to enhance model stability. Previous studies have demonstrated that UKF accounts for crossing fibers and offers sensitive and reliable fiber tracking with a wide range of populations, including variations in age, health conditions, and dMRI acquisition protocols \cite{zhang2018b,zhang2020,xue2023}. Furthermore, UKF performs simultaneous fiber tracking and multi-tensor modeling. This is more robust to partial volume effects (PVEs), allowing microstructure measures (including FA) to be directly computed at points along each fiber streamline, enabling fiber-specific microstructure estimation for accurate parcel measure computation. However, it is critical to distinguish between the pathway identified by tractography and the biological interpretation of its microstructure. For example, while we can delineate the trajectory of the CST pathway, the microstructural metrics (e.g., FA) computed along this pathway inherently reflect the average properties of the entire PyT. This is because the spatial resolution of dMRI cannot separate CST axons from other intermingled corticofugal fibers (e.g., corticobulbar fibers) within the PyT at the sub-voxel level \cite{Kjer2025}.

In our study, for each of the dMRI scans from the individual subjects, tractography parameters are adopted from previous work \cite{zhang2018b} for whole-brain WM fiber tracking. Specifically, tractography was seeded in all voxels within the brain mask where FA exceeded 0.1. Tracking was terminated when FA dropped below 0.08 or the normalized mean signal decreased below 0.06. The diffusion tensors, along with their FA values, were preserved for subsequent analyses, facilitating a comprehensive assessment of the WM microstructure.

\subsection{Compared Methods}
\label{subsec32}
In our study, we compare our method with two state-of-the-art tractometry methods, including: AFQ and BUAN. We note that AFQ and BUAN have their own pipelines for fiber tract extraction and cleaning. However, to minimize potential biases arising from inconsistent preprocessing steps across the different methods, we choose to perform quantitative analysis directly on the subject-specific fiber tracts extracted using the ORG atlas. In brief, for AFQ, all tract streamlines are resampled to 100 points, with each point’s parcel determined by its index along the streamline. The FA values of all points within each along-tract parcel are weighted based on their proposed Mahalanobis distances from the corresponding centerline node, and the obtained weighted average FA value is then assigned to the parcel as its quantitative measure. Then parcel-level group comparison is performed using a Student's test followed by FDR for multiple comparison correction. For BUAN, it is a more advanced approach to assign each streamline point to the closest point on a tract-centerline composed of 100 points, forming parcels consisting of multiple streamline points. Instead of using an overall value to represent each parcel, the FA values of all points within parcels are incorporated into a LMM for inter-group significance analysis. Then, FDR is used for multiple comparison correction across all parcel-level statistical comparisons. The AGFS-Tractometry leverages the ORG atlas for anatomically informed tract parcellation and integrates CP algorithm with a nonparametric permutation test to identify statistically significant WM communities while correcting for multiple comparisons.

\subsection{Synthetic Data Analysis}
\label{subsec33}
\subsubsection{Synthetic Dataset Creation}
\label{subsubsec331}
The goal of this experiment is to assess whether the proposed AGFS-Tractometry can successfully identify local WM regions with true group differences. To do so, we create realistic synthetic datasets to assess the ability of AGFS-Tractometry and compare methods to identify group differences in different locations and at different size scales. We leverage a subset of the HCP-YA dataset including the fiber tracts from 46 male healthy subjects, which we synthetically modify to create the two synthetic groups with a simulated true group difference. We create synthetic data of three fiber tracts, i.e., the AF, CST, and corpus callosum 2 (CC2). By applying synthetic changes at three different locations and at three different size scales per tract, we enable a comprehensive evaluation experiment.

To create synthetic fiber tract data, we first add white Gaussian noise to the FA values of all fiber points in each subject. Then we divide the subjects into two groups, G1 and G2. For group G2, we select a random location within the fiber tract to introduce a synthetic difference. This location is defined in the common coordinate system where all subject tracts are located (the ORG atlas space in this work). We identify a spherical region of interest (ROI) centered at that location with a radius r. We introduce a true group difference by adding synthetic feature changes to the fiber points belonging to the ROI in G2. This is done by increasing the group mean FA values in G2 by multiplying the FA values within the ROI by 1.5. Therefore, the ROI is the ground truth region to be detected.

To evaluate the ability to detect local regions with different sizes, for each tract, we generate synthetic data as above using ROI radii~$r$ of 12, 15, and 18\,mm. For each synthetic tract dataset, we then test the null hypothesis ($H_0: \mu_{G1}(\mathrm{FA}) = \mu_{G2}(\mathrm{FA})$).

\subsubsection{Experimental Evaluation}
\label{subsubsec332}
Given the created synthetic datasets, we perform tractometry analyses using three comparison methods to detect the ground truth regions, including: AFQ, BUAN, and our AGFS-Tractometry. For each method, we perform univariate statistical analysis for parcel-level statistics computation, followed by multiple comparison correction. All methods use the same $p$-value threshold, that is $p < 0.05$ is considered to be significantly different. For quantitative comparisons, we compute the accuracy ($\mathrm{ACC}$) of the detected region against the ground truth regions, defined as follows:
\begin{equation}
  \mathrm{ACC} = \frac{\mathrm{TP} + \mathrm{TN}}{\mathrm{TP} + \mathrm{TN} + \mathrm{FP} + \mathrm{FN}}
\end{equation}
where $\mathrm{TP}$ (true positives) are points correctly predicted to be inside the ground-truth ROI, $\mathrm{TN}$ (true negatives) are points correctly predicted to be outside the ROI, $\mathrm{FP}$ (false positives) are points incorrectly predicted to be inside the ROI, and $\mathrm{FN}$ (false negatives) are points incorrectly predicted to be outside the ROI. The average $\mathrm{ACC}$ across all ROIs and radii is reported.

\subsubsection{Experimental Results}
\label{subsubsec333}
\begin{figure}[!t]
\centering
\includegraphics[width=13.5cm,height=8.6cm]{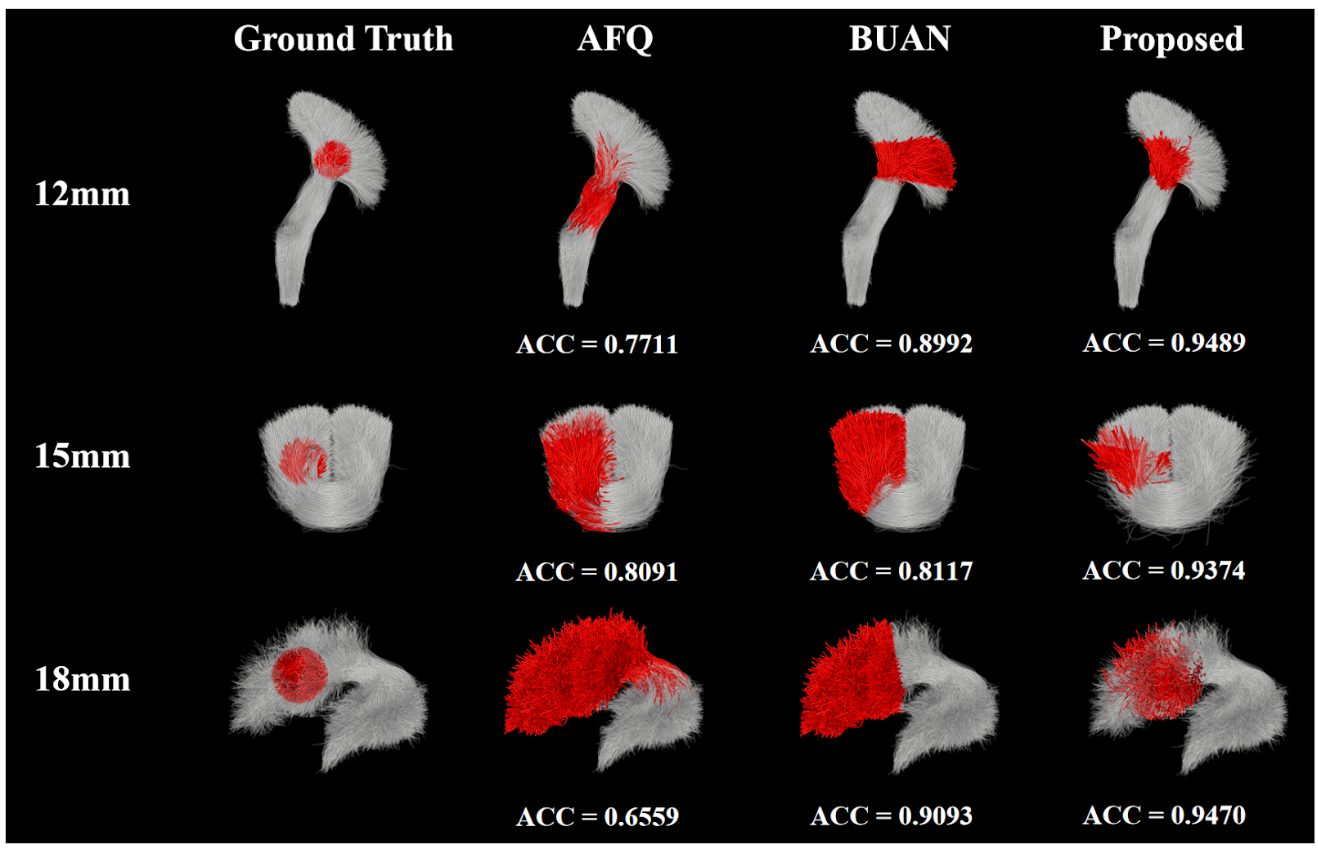}
\caption{Visual comparison across the compared methods in the synthetic datasets. Three examples out of a total of 27 combinations of different ROIs and radii across the AF, CST, and CC2 tracts are shown. The first column shows the ground truth ROIs with artificially introduced differences, while the subsequent columns show the detected regions by AFQ, BUAN, and AFGS-tractometry. For each ROI, the ACC of the detected significant region is displayed.}\label{fig5}
\end{figure}

Table \ref{tab1} gives the results of the three comparison experiments in the synthetic data experiments. By comparing performance across different tracts with specific ROIs, we can see that our proposed method outperforms AFQ and BUAN in terms of both ACC across all ROIs, except for one ROI in the CST. Figure \ref{fig5} provides a visual comparison of detection performance across three methods. AFQ tends to overestimate, extending beyond the ground truth ROI, whereas BUAN can capture more localized differences but includes false positives in adjacent areas. In contrast, AGFS-Tractometry identifies a more spatially confined and anatomically focused region of alteration, closely matching the ground-truth ROI, with consistently higher average accuracy ACC.

\begin{table}[!t]
    \centering
    \fontsize{9}{10}\selectfont 
    \renewcommand{\arraystretch}{1.2}
    \setlength{\tabcolsep}{3pt} 
    \begin{tabular}{|l|c|c|c|c|c|c|c|c|c|c|}
        \hline
        \textbf{Tract} & \textbf{Method} & \multicolumn{3}{c|}{\textbf{ROI 1}} & \multicolumn{3}{c|}{\textbf{ROI 2}} & \multicolumn{3}{c|}{\textbf{ROI 3}} \\
        \cline{3-11}
        & & 12 mm & 15 mm & 18 mm & 12 mm & 15 mm & 18 mm & 12 mm & 15 mm & 18 mm \\
        \hline
        \multirow{3}{*}{\textbf{AF}} 
            & AFQ        & 0.877 & 0.874 & 0.861 & 0.773 & 0.763 & 0.790 & 0.709 & 0.600 & 0.656 \\
            & BUAN       & 0.892 & 0.896 & 0.906 & 0.940 & 0.904 & 0.924 & 0.879 & 0.858 & 0.909 \\
            & Proposed   & \textbf{0.937} & \textbf{0.932} & \textbf{0.938} & \textbf{0.962} & \textbf{0.956} & \textbf{0.958} & \textbf{0.944} & \textbf{0.943} & \textbf{0.947} \\
        \hline
        \multirow{3}{*}{\textbf{CST}} 
            & AFQ        & 0.792 & 0.727 & 0.698 & 0.825 & 0.845 & 0.864 & 0.771 & 0.713 & 0.686 \\
            & BUAN       & 0.875 & 0.890 & 0.931 & 0.944 & \textbf{0.971} & 0.965 & 0.899 & 0.890 & 0.923 \\
            & Proposed   & \textbf{0.957} & \textbf{0.955} & \textbf{0.947} & \textbf{0.947} & 0.963 & \textbf{0.972} & \textbf{0.949} & \textbf{0.941} & \textbf{0.936} \\
        \hline
        \multirow{3}{*}{\textbf{CC2}} 
            & AFQ        & 0.809 & 0.794 & 0.793 & 0.772 & 0.733 & 0.718 & 0.760 & 0.708 & 0.659 \\
            & BUAN       & 0.812 & 0.828 & 0.790 & 0.890 & 0.855 & 0.858 & 0.841 & 0.841 & 0.801 \\
            & Proposed   & \textbf{0.937} & \textbf{0.922} & \textbf{0.903} & \textbf{0.908} & \textbf{0.921} & \textbf{0.914} & \textbf{0.937} & \textbf{0.923} & \textbf{0.924} \\
        \hline
    \end{tabular}
    \caption{Comparison results (accuracy) of AFQ, BUAN, and the proposed AGFS-Tractometry for three example fiber tracts}\label{tab1}
\end{table}

\subsection{Real Data Analysis}
\label{subsec34}
In the interest of the application to study inter-group significant differences in brain WM across populations, we apply our proposed method to two different datasets: 1) the HCP-YA dataset and 2) the ABIDE II-NYU dataset. The HCP dataset includes 100 healthy adult subjects, consisting of 54 females and 46 males. We conduct a study to investigate sex differences in the brain WM using this dataset. Additionally, to assess the reproducibility of the identified sex differences, we performed a test-retest replication analysis using another 100 subjects from the same HCP-YA dataset. This replication cohort maintained the same sex composition (54 females and 46 males) and showed no significant difference in age distribution. The ABIDE dataset includes 48 individuals with ASD and 22 HC individuals. This dataset is used to study ASD-related WM differences compared to HC.

\subsubsection{Sex-related WM Difference Using the HCP-YA Dataset}
\label{subsubsec341}
There is an evident prevalence of sex-specific differences in the brain WM, which has greatly advanced our understanding of behavioral variations between females and males  \cite{devries2009,gong2011}. In particular, language-related aspects are regarded as a critical factor in explaining sex-based behavioral differences \cite{sato2020,xu2020,gordon1991}, with extensive research demonstrating their strong association with brain WM \cite{paus2010,muer2024}. The AF is primarily associated with language functions, connecting the frontal and temporal lobes, and plays a crucial role in language processing mechanisms, such as speech production, auditory comprehension, and verbal memory \cite{catani2007,catani2009}. It links Broca’s area to Wernicke’s area, facilitating communication between regions involved in language comprehension and production \cite{musso2003}. In the interest of our study, we apply the three compared tractometry techniques (AFQ, BUAN, and our AGFS-Tractometry) to assess their ability in detecting sex-specific differences in local regions of the AF tract. For AFQ, we adhered to the standard practice of using Student's t-test, as preliminary analyses indicated that applying a LMM was not feasible due to insufficient data points per subject for reliable random effects estimation.

\begin{figure}[!t]
\centering
\includegraphics[width=13cm,height=6.5cm]{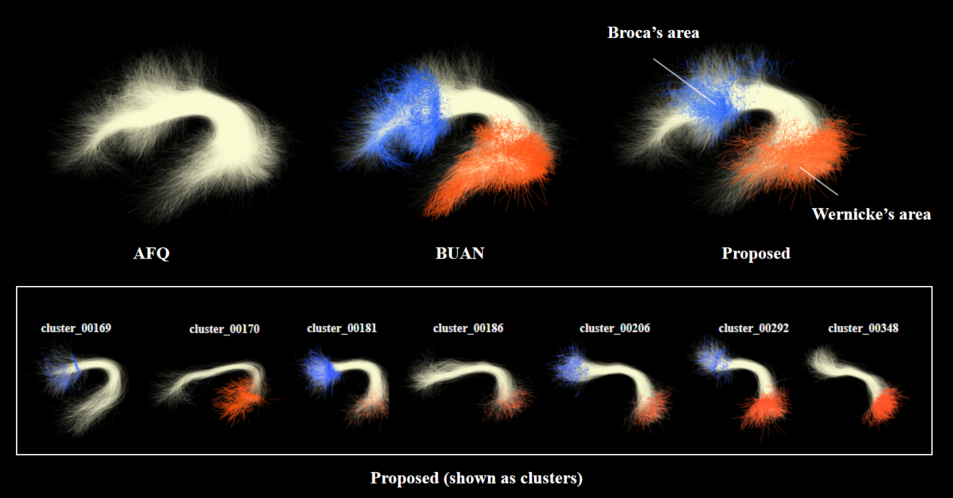}
\caption{Comparison of three methods in detecting sex-related differences in the left AF using HCP-YA data. Colored regions indicate parcels with significant group-level WM differences.}\label{fig6}
\end{figure}

Figure \ref{fig6} gives the comparison results in detecting sex-related WM differences along the left AF using HCP-YA data. Although ground truth is unknown, our method and BUAN identified partially overlapping regions of potential difference, while AFQ did not detect any significant regions. The detection of smaller, more focal regions by our method may reflect higher spatial specificity. Prior studies have reported sex-related differences in the left AF, which connects language-related areas such as Broca's and Wernicke's regions \cite{catani2005,rojkova2016,keller2009,kanaan2012}.

To further validate the robustness and reproducibility of our findings, we conducted two additional experiments: (1) a test-retest reliability analysis and (2) an independent cohort validation. The details are as follows:

\textbf{Test-Retest Reliability}: We evaluated the test-retest reliability of our proposed AGFS-Tractometry method using the HCP test-retest dataset (N=36 subjects). For the 11 clusters identified along the left AF, we computed the intraclass correlation coefficient (ICC) between the test and retest scans for each corresponding parcel. The results demonstrated good to excellent reliability across all clusters, with 4 clusters exhibiting ICC values above 0.75 and the remaining 7 clusters falling in the range of 0.60-0.74. According to the established guidelines by (Cicchetti 1994), this indicates a high degree of measurement consistency. Furthermore, a direct group comparison between the test and retest data using our method revealed no statistically significant differences in the left AF, confirming the high reproducibility of our measurements. 

\textbf{Independent Cohort Validation}: To demonstrate the generalizability of our initial sex-difference findings, we replicated the analysis on an independent, matched cohort of 100 new subjects from the HCP dataset. This new cohort was selected to have an identical sex distribution and a comparable age range (with no significant difference in age) to the 100 original subjects. Applying our method to this validation cohort, we successfully identified localized sex-related differences in the left AF. The spatial pattern of these significant regions was highly consistent with our original results in Figure 6. Specifically, when the significant regions from both analyses were binarized and compared as volumetric masks, they exhibited an 84.24\% spatial overlap, predominantly located near Broca's and Wernicke's areas. This strong concordance across independent datasets underscores the robustness and reliability of our proposed tractometry framework.

\subsubsection{ASD-related WM Differences Using the ABIDE Dataset}
\label{subsubsec342}
Previous studies have indicated that ASD is associated with alterations in several tracts, which are crucial for interhemispheric communication, cognitive functions, and social regulation. Given the widespread distribution of affected tracts reported in the literature \cite{ecker2012,hoppenbrouwers2014,zhang2018}, our analysis was not restricted to a small set of predefined tracts but instead included all major tracts represented in the ORG-atlas. This allowed us to assess the potential of each method in detecting ASD-related WM differences across a broad anatomical range, without implying a whole-brain multiple comparison correction.

\begin{figure}[!t]
\centering
\includegraphics[width=13cm,height=7.5cm]{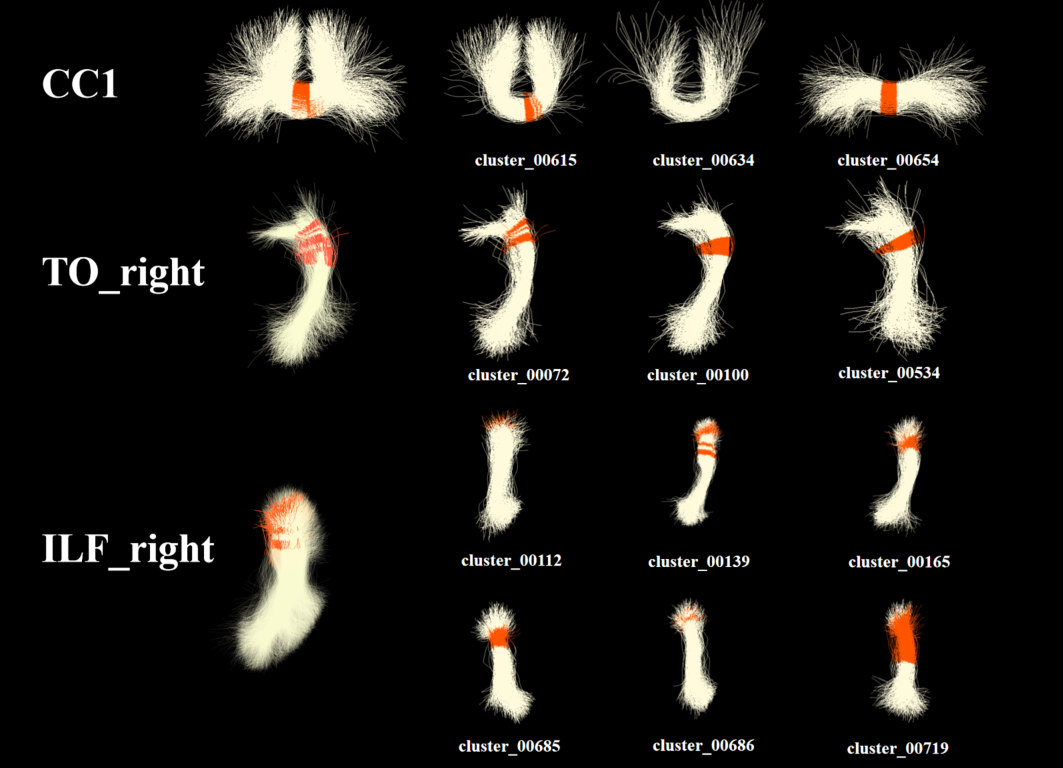}
\caption{Application to real data (ABIDE) to evaluate the ability of different methods to identify potential group-level WM differences (colored regions). The proposed method identified localized differences in three tracts (CC1, ILF\_right, TO\_right), whereas AFQ and BUAN did not. These results illustrate the comparative sensitivity of the methods, rather than aiming to establish definitive ASD-related findings.}\label{fig7}
\end{figure}

We performed AFQ, BUAN, and our method on the 58 major tracts of all subjects in the ABIDE dataset to explore group-level differences and assess the ability of each method to localize potentially altered regions along the tracts. While neither AFQ nor BUAN detected any statistically significant differences, our method identified several regions showing significant group differences in the corpus callosum 1 (CC1), right inferior longitudinal fasciculus (ILF\_right), and right temporo-occipital tract (TO\_right), as shown in Figure \ref{fig7}. These tracts have been reported to be previously associated with ASD-related alterations \cite{yoon2022,li2022,bernal2009}, suggesting that our method may better capture subtle group-level differences in brain WM, although the true underlying differences remain unknown in this real-world dataset.

\section{Discussion}
\label{sec4}
In this paper, we propose a novel atlas-guided fine-scale tractometry methodology that leverages tract spatial information and permutation testing to enhance the along-tract statistical analysis between populations. We evaluate the performance of methods by applying them to two kinds of datasets: (1) the synthetic dataset with artificially created ground truth for quantitative evaluation of detection performance (e.g., accuracy), and (2) two real-world datasets used for qualitative assessment of methods’ ability to identify plausible group-level differences. In the synthetic dataset with known inter-group differences, results showed that AGFS-Tractometry was more accurate at detecting ground truth differences than the popular tractometry techniques AFQ and BUAN.  In the real data analysis experiments, AGFS-Tractometry can identify more regions with significant differences, suggesting its potential to detect subtle or spatially localized WM group-level differences. Several key observations about the results are discussed below. 

We propose a novel tractometry framework that utilizes the anatomical guidance of the ORG-atlas to perform a fine-scale, along-tract parcellation for tract profiling. By assigning streamlines to anatomically informed WM clusters, we can assess the underlying structural subdivisions of each tract. This is an improvement over the existing methods that are performed on the entire WM tract, without accounting for its finer substructures to accurately capturing the fine-scale structural organization of WM. For example, AFQ adopts an ordinal parcellation strategy along streamlines, which assumes consistent streamline lengths and alignment across subjects. This assumption may be inaccurate in tracts with high inter-subject variability, potentially leading to less precise regional attribution. BUAN, on the other hand, relies on a single centerline-based approach to profile the tract, which can result in irregular segment orientations, particularly near tract endpoints where streamlines tend to fan out. Numerous studies have demonstrated that fiber tracts contain finer subdivisions with distinct microstructural and functional properties \cite{he2023,bernal2009,rilling2008}. As a result, AFQ and BUAN may inadvertently mix different anatomical regions into a single parcel, introducing structural and functional heterogeneity that may pose challenges for detecting localized group-level WM differences. 

However, our framework is not limited to the ORG-atlas. It can be applied to any fiber tract atlas that meets two fundamental criteria: (a) it provides sufficiently fine subdivisions along the length and across the diameter of the fiber tracts, ensuring a fine-scale parcellation; (b) it can be applied to individual subjects to identify subject-specific fiber tract subdivisions, allowing for robust group-wise statistical inference. These criteria ensure that the framework can achieve a fine-scale parcellation and establish a common space for all subjects, facilitating robust group comparisons. Furthermore, it is important to acknowledge the anatomical limitations of our proposed framework. While our method offers a substantial increase in spatial sampling resolution within a tract, the anatomical precision of the resulting parcels is inherently constrained by the accuracy of the underlying tract definitions provided by the ORG atlas. This challenge, however, is not unique to our method but is a general consideration for all tractography-based parcellation approaches. Consequently, our findings should be interpreted as demonstrating the method’s sensitivity to localized microstructural differences along the tract, whose exact anatomical correspondence warrants further validation. Furthermore, to ensure the reliability of microstructural measurements, the parcel size must be considered in relation to the image resolution limit \cite{dyrby2025}. To address this, we perform an  analysis by calculating the voxel count of each parcel occupied. For the CST, our method subdivides the entire tract into a total of 1100 parcels, each occupying an average of 41 voxels across the 100 HCP subjects. This mean parcel size substantially exceeds the practical resolution limit as pointed out by Dyrby et al \cite{dyrby2025}, confirming that our fine-scale parcellation is methodologically robust in this regard and yields microstructurally reliable measurements.

For statistical inference, we adopt a nonparametric permutation test to control for multiple comparisons. Unlike conventional corrections such as Bonferroni or FDR, which assume independence or specific distributional forms, the permutation test is assumption-light and well-suited for simultaneous analysis of multiple WM parcels. In real-world analyses, including the HCP-YA and ABIDE II datasets, AGFS-Tractometry identified more regions with statistically significant group-level differences than AFQ and BUAN, highlighting its potential for enhanced sensitivity in detecting WM alterations. In the HCP-YA dataset, our method identified sex-related differences in regions anatomically close to Wernicke’s and Broca’s areas, which are known to be involved in language processing. These findings align with prior studies indicating sex-based WM differences in these regions, potentially linked to variations in language processing and cognitive function [44,53], providing partial validation of our method’s ability to localize meaningful WM differences. In the ABIDE dataset, AGFS-Tractometry identified significant group-level differences in the ILF, TO, and CC1, all of which have been implicated in ASD-related disruptions of interhemispheric communication, visual processing, and social cognition \cite{hoppenbrouwers2014,nair2013}. In contrast, AFQ and BUAN did not yield significant findings. These results suggest that AGFS-Tractometry may have enhanced sensitivity for localizing putative group differences in WM structure. However, we emphasize that these real-data analyses serve primarily as method validation, and further studies with larger cohorts and clinical validation are required to draw definitive neurobiological conclusions.

Our findings should be interpreted considering two general challenges in tractometry. First, it is crucial to distinguish between the anatomical pathway identified by tractography and the microstructural properties inferred from it. Our streamline reconstructions delineate the CST pathway based on its trajectory from the motor cortex to the spinal cord. While this pathway is anatomically a core component of the broader PyT \cite{Thomas2014,Jones2010}, the spatial resolution of dMRI cannot isolate it from other co-localized corticofugal fibers within the PyT at a sub-voxel level \cite{Kjer2025,chenot2019,JANG201580}. Consequently, while we analyze microstructural metrics (e.g., FA and MD) along the defined CST pathway, these measurements inherently reflect the average properties of the entire PyT at each location. This distinction is vital for interpreting our tractometry findings within a biologically meaningful framework. Second, the PVEs near the gray-white matter interface can affect the reliability of diffusion metrics in terminal regions (e.g., the AF in Figure 6). While our methodology mitigates this concern through high-resolution data, a PVE-robust UKF tractography algorithm, and consistent pipeline application, the inherent reliability challenges in these regions necessitate cautious interpretation.

Potential future directions and limitations of the current work are as follows. First, our current framework evaluates a single diffusion metric at a time, which may limit the generalizability of the results across different diffusion properties. A future research direction could incorporate multiple complementary diffusion measures to improve generalizability and robustness \cite{yeatman2012,garyfallidis2014}. Second, the current framework is limited to two-group comparisons. Extending it to support multi-group or continuous-variable analyses will be important for broader applicability in neuroimaging studies. Third, while the CP algorithm provides a powerful approach to identifying spatially connected suprathreshold WM parcels, its computational cost remains a bottleneck, particularly under permutation-based inference. Optimizing this step through parallel computing or more scalable community detection strategies may improve efficiency. Lastly, although this study does not aim to make definitive neurobiological claims, our results highlight the potential of AGFS-Tractometry to support future investigations into clinically and cognitively relevant WM differences using larger, more diverse datasets. The method provides a foundation for finer-scale localization and hypothesis generation in future brain-behavior studies.

\section{Conclusion}
\label{sec5}
We propose AGFS-Tractometry, a novel atlas-based method for fine-scale along-tract WM analysis. It leverages anatomical guidance for tract parcellation and integrates local neighborhood information to improve spatial coherence. Compared to AFQ and BUAN, which rely on positional indices or centerline profiling, our method offers a complementary approach with enhanced spatial resolution. We apply a nonparametric permutation test for group comparison, with correction for multiple comparisons across parcels. While AGFS-Tractometry identifies more localized differences in both synthetic and real datasets, these findings are intended to evaluate methodological performance rather than draw biological conclusions. Larger datasets and more advanced statistical models are needed for future clinical applications. Overall, AGFS-Tractometry complements existing tools and may help support finer-scale WM analysis in future neuroimaging studies.

\section{Data and Code Availability Statement}
\label{sec6}
The code used in this study, along with the tract profiling template, is publicly available at https://github.com/ZhengRuixi/AGFS-Tractometry.git. A license has been included in the repository to facilitate use by other researchers.

\section{Acknowledge}
\label{sec7}
This work is in part supported by the National Key R\&D Program of China (No. 2023YFE0118600) and the National Natural Science Foundation of China (No. 62371107).

\bibliographystyle{elsarticle-num} 
\bibliography{ref}       
\end{document}